# ClassSpy: Java Object Pattern Visualization Tool

Tufail Muhammad, Zahid Halim
Faculty of Computer Science and Engineering
Ghulam Ishaq Khan Institute of Engineering Sciences and
Technology, Topi, Pakistan
tufail@giki.edu.pk, zahid.halim@giki.edu.pk

MajidAli Khan
College of Information Technology
Prince Mohammad Bin Fahd University
KSA
majidkk@gmail.com

*Abstract*—**Modern java programs consist of large number of classes as well as vast amount of objects instantiated during program execution. Software developers are always keen to know the number of objects created for each class. This information is helpful for a developer in understanding the packages/classes of a program and optimizing their code. However, understanding such a vast amount of information is not a trivial task. Visualization helps to depict this information on a single screen and to comprehend it efficiently. This paper presents a visualization approach that depicts information about all the objects instantiated during the program execution. The proposed technique is more space efficient and scalable to handle vast datasets, at the same time helpful to identify the key program components. This easy to use interface provides user an environment to glimpse the entire objects on a single screen. The proposed approach allows sorting objects at class, thread and method levels. Effectiveness and usability of the proposed approach is shown through case studies.**

## I. INTRODUCTION

Most of the modern age computer software solutions are very complex in their structure and functionalities. These software solutions consist of tens and hundreds of packages and classes having large numbers of methods or functions. With such a large number of classes and complex functions, vast amount of objects are instantiated by these programs during execution. Software developers are always enthusiastic to know about the most important classes or methods of their programs. Understanding object creation by a particular class and categorizing on the basis of package, class or method is not a trivial task for the software developers.

On the other hand, software visualization is used as an effective technique to comprehend software at various levels as well as to cope with the software complexity. Some tools and techniques are used to visualize software static aspects [1-3] others are related with dynamic analysis of a software such as tracing data [22] or component interaction at runtime of the program [6, 7]. The basic drawback of the most of software visualization tools is the task oriented aspect of these tools; that is why developers need a sophisticated tool to meet their requirements. Usually programs generate a large amount of data at runtime i.e. thousands of events for object creation; hence, analysis of this data is not an easy task. Visualization facilitates developer in program comprehension in a more efficient manner [22]. Visualization technique is more effective to convey information as compare to other techniques e.g. sonification[34][35].

This paper presents a software visualization methodology in shape of a tool that provides developers an environment to understands their program components i.e. packages, classes, methods and threads without knowing its source code. The proposed visualization is based on dynamic data extracted from a program during execution time. The tool depicts all the objects of a program created by various classes during its execution. The technique used by the tool is scalable to handle large amount of events on a single screen. The developer will be able to sort the available information on various levels. The system can be used with legacy software where source code is not available to understand it components as well as for pedagogical purpose.

The rest of this paper is organized as follows: Section 2, briefly review the related work. Section 3 introduces the new visualization techniques. Section 4 lists case studies and Section 6 concludes the paper.

## II. RELATED WORK

Software visualization is used as a technique to visualize various aspects of software at different granulate levels. The software visualization tool in contemporary research work is domain and task specific. There are many techniques and tools for task effective program comprehension and identification of various components [19, 31]. The software visualization tools presented in this work mainly covers two software aspects, static analysis [21] and dynamic analysis [22]. The visualization tools presented in [4, 6,7,15] are based upon the use of dynamic analysis to understand program's runtime behavior effectively. Other tools use software visualization for pedagogical purpose to understand program [17]. Research on grid based visualization is also presented in previous works [27]. The current techniques in component identification area mostly do not consider visualization techniques and method. The proposed visualization tool is different from all of the existing ones; our approach shows the component and objects on a single screen. The work presented by Zaidman in [19] identifies the key classes in software using dynamic coupling and webmining techniques. A novel method has been proposed by Kawrykow and Robillard in [24] that automatically detect API imitation in software code using static analysis techniques. The technique helps software developer to identify API usage and its patterns. Another approach for API usage has been adopted in [25] by Souza and

Bentolila. The work presented in [33] used clusters to identify key components in software system. Their work has two facets; initially to find usability of API using complexity metrics and then utilized visualization to present this information to the developer effectively. The previous work, discussed in this section has two main limitations, first; most of the techniques are not based on dynamic analysis and mostly concern with static aspect of software. Secondly no suitable visualization is provided in the literature to visually analyze data. We address these two limitations and our work aims to visualize dynamic data of program and provides visualization tool to assist programmer more effectively.

### III. PROPOSED VISUALIZATION

This section explains the proposed visualization system. Below is an overview of the whole system and how it works.

#### A. System Overview

The proposed visualization system's design and working can be divided in three main steps: 1) instrumentation packages preparation, 2) data collection and 3) visualization. The first step is to prepare instrumentation packages class used to insert probe into program code at runtime. These classes are important and build with care, since one wrong statement can damage the target program. Second step is to run the target program under the control of instrumentation code to collect data generated from program. The data generated from the program are stored in log file as CSV format. The information we collect from a target program includes, object name, type, package name, class and method name, data/time and line number. Visualization module is the last step where we take the log file as input and create the visualization. The visualization module is used to map data to visual elements and portray it on the screen. The proposed visualization consists of small squares depicted on screen, where each square represents an object of a particular class in the program. The color of a square shows different categories of depicted information. The user can sort these squares based on various components of a program such as package, type, method etc. The sorting options are available to user through menu. Through this software developer is able to identify key program components that are responsible for objects creation. Fig. 1 gives the system overview. The whole system is implemented in java using Eclipse IDE, while the instrumentation class is build using BCEL [32]. The following subsections explains the system components in more detail.

#### B. Instrumentation Packages

In the first phase instrumentation classes are prepared. These are used to generate and collect runtime data from a target program. We used dynamic instrumentation, where class bytecode is instrumented at load time of a particular class. Bytecode Engineering Library (BCEL) [32] which is an open source library from Apache software is used to build instrumentation classes. BCEL provide Java Virtual Machine (JVM) independent environment for instrumentation with flexible control. The instrumentation code, consists of two packages, first package contains those classes responsible for adding probe to specific position in the target program bytecode and the actual code to be inserted. The second package is just like utility code that monitors the interesting events of the program and records them in a log file. The dynamic instrumentation process is tedious task as the volume of data generated is massive. We have used selective instrumentation to cope with log file size and only those methods are instrumented where objects are instantiated, deleted or method entry events occurs.

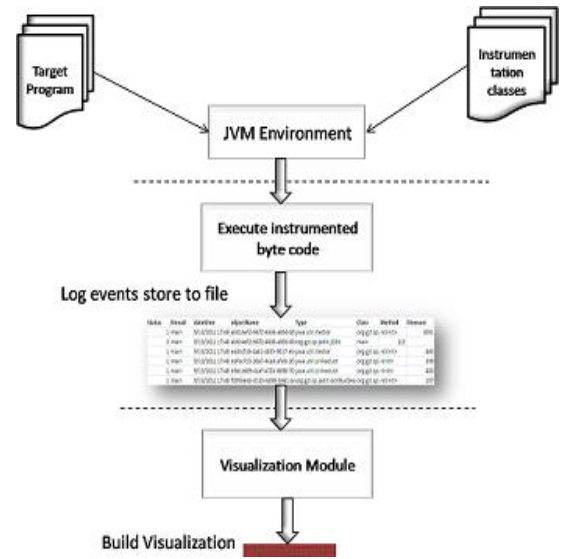

Fig. 1. System Overview

#### C. Data Collection

As the instrumentation packages are built, next step is to take a target program and run it under the control of instrumentation code. As mentioned previously, interesting events have been recorded to log file. For this purpose we collect specific information from the program during execution time without knowing its source code. The data extracted in this case includes: 1) object unique identifier, 2) object owner thread, 3) object type, 4) time of event and 5) object status. It may be noted that some of these information can be collected at static level, in the case where source code is available. Rest of the data e.g. source code line number, package and class name, are static features. These are also gathered at execution time. Fig. 2 shows a snapshot of log file created during our experiment. The data collected is for short time durations. The events may be recorded to log file as long as the program runs, this is constraint to space available for storing log file. We can also record various scenario based information, so the developer will identify the key elements during the processing of these scenarios.

Fig. 2. Segment of log file

*D. Visualization and User Interaction*

The last step in this process is visualization and providing an effective user interface facility. Given an input file from the previous step, visualization module maps the data elements of input file to visual elements and creates grid-based visualization. As described previously, input file consists of individual record for each object created. Visualization module maps each object to square cell on to the grid. The number of cells and dominion of each cell depends on the size of presented input file. The main objective of visualization is to provide the developer, a view to identify main class and how many objects are created by them. Developer then would be able to sort the information on various program components i.e. packages, methods and threads. The visualization will help the developer in better understanding of the program, without seeing source code. The visualization module may extend for other usage situations as well. The algorithm for visualization is listed in Fig 3.

*1) Visual Encoding and Layout:*

As described earlier, the input file contains all the information about each object; we build visualization on the basis of this information and relationship. Since we have large number of events in the log file, for the visualization we need a space efficient and scalable technique that is able to show all the information on a single screen. Proposed layout algorithm first determines the number of objects to be shown and then distributes the screen area among all the objects. The dimension of each cell in the grid can vary with the number of records in input file. As the number of objects grows in input file the square size gets smaller. The layout algorithm first takes the collection of square and displays each square row wise on the screen. Each time the user sorts the screen information on some other attribute, the layout algorithm is executed and it repaints the entire visualization in respective order.

The color of each square is computed from each object's properties using hash function, which gives a unique color for each value. We have used a simple RBG color scheme. The objects under same property are assigned same color, which

---

**Input:** *A log file, consist of events of objects creation, in CSV format*
**Output:** *The Visualization*
1  F ← {all records store in log file}
2  While End of F
3     Token ← {each column of record of F}
4     Array ← {store each Token}
5  While END of Array
6     ObjectArray ← {Array element}
7     Compute screen dimension
8     For each objectArray Element
9        Draw color square on screen

---

Fig. 3. Visualization Algorithm

allow developer to recognize patterns in a single glance. As shown in Fig. 4 squares are depicting objects of different types shown with various colors. If developer changes the sorting order, the squares are sorted accordingly and the color for each square is recomputed using hash function taking respective attribute as input. The larger the numbers of squares with a particular color means that there are more objects of same class/type.

*2) Interaction:*

The proposed visualization supports user interaction through the interface of the tool. The interaction enables developer to find as well as sort the available information in different ways. Initially developer sees the visualization based on class level information, further this can be changed to view accordingly, through the menu interaction provided. User is also provided with optional menu to load different files to the visualization scenario. Menu options are available to user for sorting the information on the basis of package, type, thread or method. If user sorts the information on the bases of type; the square representing objects of same type are assigned same color. Similarly if user sorts the same information on thread, the tool gets the number of threads in the program as well the number of objects handled by each thread. User can view information about a particular object by clicking the respective square; a tooltip appears displaying information. We may further extend our prototype model to add search based facility, through which user would find more specific information.

IV. CASE STUDIES AND DISCUSSIONS

This section presents case studies to evaluate tool usability and effectiveness from developer's perspective. We illustrate that the tool helps in program comprehension and in identifying program package/classes/method with more effectiveness. We take jEdit [20] as a target program and extract runtime information with our instrumentation code. jEdit is java based editor for programmer consisting a large number of classes and packages and simple short time run of program produces a vast amount of data. We run jEdit on a 2.93GHz Intel corei3 system for few minutes, where our instrumentation package comes to work and instrument jEdit classes, before the actual execution. The events monitoring and recording can be carried out for long duration provided large storage space is available. The interesting events are monitored and all information is logged to a CSV file. Visualization main view available to the developer is shown is Fig. 5. Here the objects of each type are shown in the order they are recorded to input file. Each distinct color shows different type of object creation during program execution. The various classes found in this data are shown in the table, where the class with name org.gjt.sp.jedit.GUIUtilities creates large number of objects. The visualization view may be sorted according to thread attribute of the object, and we have a view as shown in Fig. 6.

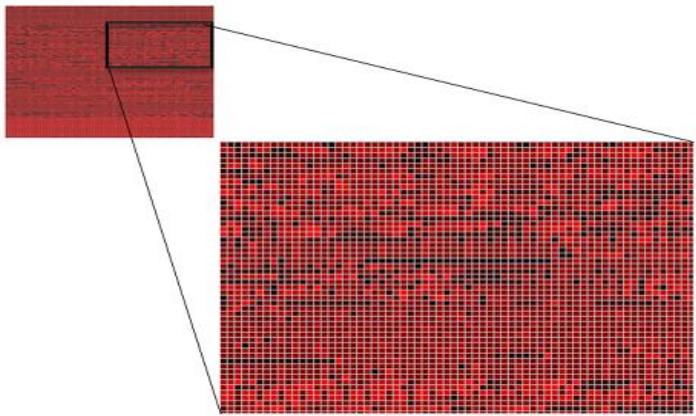

Fig 4. Zoom view

It is found from the depicted information that there are three threads responsible for handling program objects i.e. main, Thread-0 and AWT-EventQueue-0, but still most of the program objects are created under main thread and AWT-EventQueue-0, while the Thread-0 is responsible for destroying objects. One reason for this may be that it has happened due to the beginning of the program; later on some other threads were created. Similarly, the visualization can be sorted according to various levels and perspectives available from the menu. All this information through visualization reveals all the knowledge about the program components like, threads, classes, packages, methods without knowing the actual program code. The developer starts from here to understand program and its building blocks.

## V. CONCLUSION AND FUTURE WORK

This paper presents a grid-based visualization tool to support java program understanding through identifying key classes/packages. The visualization is build from data extracted at runtime through the instrumentation process. The tool provides a simple interactive environment to assist developers in evaluating java program without witnessing the source code and can be used for pedagogical purposes as well. The tool provides a space efficient and scalable approach that handles large amount of data displayed on a single screen. Proposed tool would be further improved to show class functionality and interaction with other classes, at the same time the visualization techniques may be used in other domains for information clustering.

ACKNOWLEDGMENT

This research has been supported by the Higher Education Commission (HEC) of Pakistan under the Indigenous Fellowship Program. The authors would like to thank GIK Institute of Engineering Sciences and Technology, for providing research facilities.

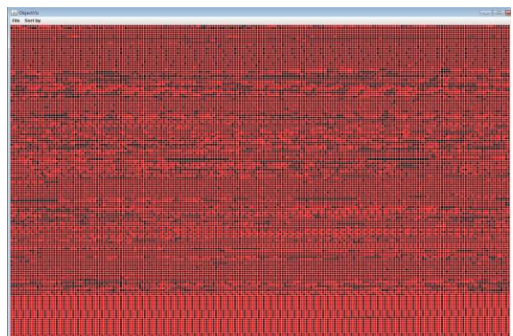

(a)

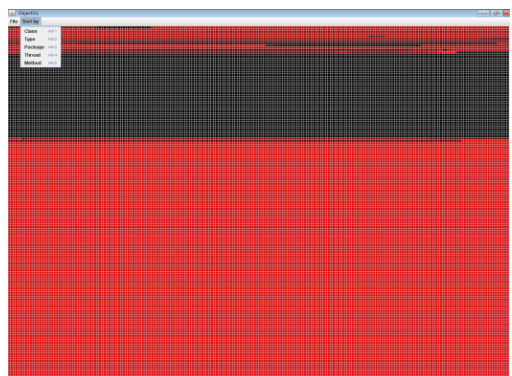

(b)

Fig. 5. View (a) Unsorted (b) Sorted by type

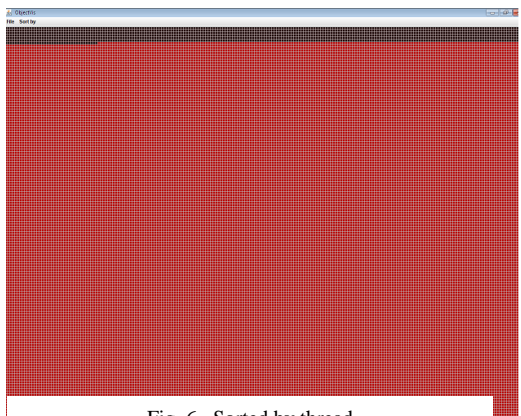

Fig. 6. Sorted by thread